\documentclass[aps, prx, twocolumn, notitlepage, longbibliography]{revtex4-1}
\usepackage[left=1in,right=1in,top=1in, bottom=1in]{geometry}
\usepackage{amsfonts}
\usepackage{graphicx}
\usepackage{wrapfig}

\usepackage{enumerate}
\usepackage{color}
\usepackage{times}
\usepackage{titlesec}
\usepackage[normalem]{ulem}
\usepackage{textcase}
\usepackage{url}
\usepackage{natbib}
\usepackage{hyperref}
\usepackage{amsmath,bm}

\newcommand{\bea}{\begin{eqnarray}}
\newcommand{\eea}{\end{eqnarray}}
\newcommand{\be}{\begin{eqnarray}}
\newcommand{\ee}{\end{eqnarray}}
\newcommand{\bw}{\begin{widetext}}
\newcommand{\ew}{\end{widetext}}

\begin{document}

\title{General theory of many body localized systems coupled to baths}
\author{Rahul Nandkishore}
\affiliation{Department of Physics and Center for Theory of Quantum Matter, University of Colorado at Boulder, Boulder, CO 80309, USA}
\author{Sarang Gopalakrishnan}
\affiliation{Department of Physics and Burke Institute, California Institute of Technology, Pasadena, CA 91125, USA }
\begin{abstract}
We consider what happens when a many body localized system is coupled to a heat bath. Unlike previous works, we do not restrict ourselves to the limit where the bath is large and effectively Markovian, nor to the limit where back action on the bath is negligible. We identify limits where the effect of the bath can be captured by classical noise, and limits where it cannot. We also identify limits in which the bath delocalizes the system, as well as limits in which the system localizes the bath. Using general arguments and dimensional analysis, we constrain the overall phase diagram of the coupled system and bath. Our analysis incorporates all the previously discussed regimes, and also uncovers a new intrinsically quantum regime that has not hitherto been discussed. We discuss baths that are themselves near a localization transition, or are strongly disordered but 
protected against localization by symmetry or topology. We also discuss situations where the system and bath have different dimensionality (the case of `boundary MBL' and `boundary baths').
\end{abstract}
\maketitle

\section{Introduction}
The idea of localization dates back to Anderson's seminal work in 1958 \cite{Anderson1958}. 
While Anderson's discussion was general, the phenomenon was nonetheless widely assumed to be particular to systems of non-interacting particles (but see Refs.~\cite{Fleishman, MaksimovKagan}). 
A decade ago strong perturbative arguments were put forward \cite{AGKL, Mirlin, BAA} in support of the idea that interacting many body quantum systems could be in a localized phase, where they failed to equilibrate even at infinite times - a phenomenon that was dubbed `many body localization' (MBL).  Numerical works \cite{Prosen, OganesyanHuse, PalHuse} and a recent mathematical proof \cite{Imbrie} have not only put this {\it many body localized} phase on firm ground, but have established that MBL can even persist into the regimes of strong interactions and high energy densities that are inaccessible to perturbation theory. More recently, it has been realized \cite{LPQO, VoskAltman2014, Pekkeretal2014, Bahrietal2013, Chandranetal2014} that MBL can support exotic forms of quantum order at high energy densities, even when such types of ordering are forbidden in thermal equilibrium. MBL is also associated with a rich phenomenology, including an emergent integrability \cite{HNO, Serbynlbits}, an unusual pattern of entanglement \cite{geraedts}, a nonlocal response to local perturabtions \cite{nonlocal} and novel behavior in linear response \cite{gopalakrishnan}. For a review, see \cite{ARCMP}. For all these reasons and more, MBL has excited tremendous interest, which has only heightened since the phenomenon was potentially observed in ultracold atoms experiments \cite{bloch1, bordia, bloch2}. However, despite the enormous interest in MBL, most investigations of this phenomenon have focused on {\it closed} quantum systems, perfectly isolated from any environment. Any `realistic' experimental system will always be coupled, however weakly, to a thermalizing environment. Additionally, in a great many settings, including systems with protected delocalized states \cite{QHMBL, BanerjeeAltman, spinbath} and continuum systems \cite{Aleiner, 2dcontinuum} an `internal' heat bath may also be inevitably present in the system. How then should one understand MBL systems coupled to baths?

In a series of works \cite{QHMBL, Aleiner2, BanerjeeAltman, Aleiner, 2dcontinuum, NGH, GN, JNB, HNPRS, proximity, Hyatt, Fischer, spinbath, lesanovsky2016, everest2016role, Znidaric}, a partial answer to the above question has emerged. In \cite{NGH}, the `generic' case of an MBL system coupled to a `good' bath was considered (this situation was also studied numerically in \cite{JNB}). It was pointed out that while weak coupling to a bath causes the {\it eigenstates} to become effectively thermal, nonetheless signatures of MBL survive in the {\it dynamics} (as characterized by spectral functions) as long as the coupling is weaker than the characteristic energy scales in the system. A logarithmic enhancement of the relaxation rate particular to MBL systems was also identified, as were various experimental diagnostics of MBL. Recently, this situation has also been analyzed within the Lindblad formalism in \cite{Fischer, Znidaric, lesanovsky2016, everest2016role}. In \cite{GN}, a {\it narrow bandwidth} bath (able to supply only small amounts of energy) was considered, and it was pointed out that the relaxation rate should have an additional power law smallness in the bandwidth of the bath. Additionally, this perspective was used to develop a self consistent mean field theory of the many body localization transition. All of the above works stayed in the regime where the `back action' of the system on the bath was negligible. In \cite{HNPRS}, single particle localized systems coupled to baths were investigated, including in the regime of strong back action. In \cite{proximity}, it was pointed out that in this strong back action regime, the system could localize the bath instead of the bath delocalizing the system, a phenomenon that was dubbed the `MBL proximity effect.' This scenario was investigated numerically in \cite{Hyatt}. 

There has thus emerged a large body of work examining MBL systems coupled to baths. However, each of these works operates in a particular corner of parameter space, and the full spectrum of possibilities has never been organized or systematically surveyed. Additionally, all of the above works have focused on situations where the system and bath are coupled {\it everywhere}, ignoring the interesting issue of a system and bath coupled only on the {\it boundary} (a partial discussion of this `codimension one' problem was recently provided in \cite{Chandran}). Nor has it been clarified under what circumstances the bath may be modeled as a source of classical noise, 
as is done when using the Lindblad formalism with dephasing noise, as in Ref.~\cite{Fischer, Znidaric, lesanovsky2016, everest2016role}. 

In this paper we provide a general theory of MBL systems coupled to baths. We begin in Sec.\ref{models} by introducing the basic models we will use to illustrate our discussion. In Sec.\ref{noise} we discuss MBL systems coupled to {\it classical} stochastic noise, where the noise serves as a minimal model for the effect of a bath. We discuss the limits in which classical noise is a good model for the bath, and when quantum effects become important. In Sec.\ref{mblbath} we consider MBL systems coupled to thermalizing quantum systems in the traditional `co-dimension zero' setting where the system and bath are of the same dimensionality and are coupled everywhere, and where the bath can be characterized by a single timescale. We point out that the behavior of the resulting coupled systems is governed by a small number of parameters. We organize the parameter space in terms of dimensionless ratios of parameters, and point out that the previous works \cite{NGH, GN, JNB, HNPRS, proximity, Hyatt, Fischer} can all be understood as describing particular corners of parameter space. In Sec.\ref{specialcases} we discuss the `spectral diffusion' scenario, where the correlation time in the bath is much longer than the inverse bandwidth. This situation obtains when the bath is close to a localization transition, or for situations where a bath is strongly coupled to the system but `protected' against localization by symmetry, topology, or long range interactions. We also identify a regime that had not previously been explored, and discuss the behavior therein. In Sec.\ref{boundary} we turn our attention to the `co-dimension one' case. We discuss the phenomenology of MBL systems with `boundary baths' and thermal systems with MBL layers deposited on the boundary, paying particular attention to counterintuitive near boundary phenomena that can arise in certain regimes. We conclude in Sec.\ref{conclusions} by summarizing the general principles guiding our understanding of MBL systems coupled to baths. The appendix provides technical details referred to in the main text, and also discusses the special case of non-interacting systems and baths. 

\section{Model}
\label{models}
The system we consider is the following: a system on a $d$ dimensional lattice with two species of particles - A and B. The particles can be spinless fermions for specificity, although we are working at high temperatures where the particle statistics are likely unimportant. The A particles are present with density $n_A$ and have Hamiltonian 
\begin{equation}
H_A = \sum_{\langle ij \rangle} t_A c^{\dag}_{i} c_{j} + U_A c^{\dag}_i c_i c^{\dag}_j c_j + \sum_i \epsilon^A_i c^{\dag}_i c_i 
\end{equation}
where $t^A$ is the hopping, $U^A$ is a nearest neighbor interaction, and $\epsilon^A$ is a random potential, drawn from a distribution of width $\mathcal{W}$. The width of the distribution is sufficiently large that the A particles in isolation are in an MBL phase, with a localization length $\xi_A$ and an associated energy scale $ W = \mathcal{W} \exp(-s \xi_A^{d})$, where $s$ is the entropy density. We do assume we are working well away from the trivial limit $t = 0$, and also well away from the non-interacting limit, so that the `many body level spacing' is the only relevant quantity.  Meanwhile, the B particles  have $N$ flavors, are present with  total density $n_B \approx N n_A$, and have Hamiltonian 
\begin{eqnarray}
H_B &=& \sum_{\langle ij \rangle, \alpha } t_B d^{\dag}_{i, \alpha} d_{j, \alpha}  + \sum_{i, \alpha} \epsilon^B_i d^{\dag}_{i,\alpha} d_{i,\alpha}  \\&+& 
 \sum_{i \alpha \beta} U_B d^{\dag}_{i, \alpha} d_{i, \beta} d^{\dag}_{i, \beta} d_{i, \alpha}  +\sum_{\langle ij \rangle, \alpha \beta} U_B' d^{\dag}_{i, \alpha} d_{i, \beta} d^{\dag}_{j, \beta} d_{j, \alpha} \nonumber
\end{eqnarray}
where $\alpha$ and $\beta$ are flavor labels, $i, j$ label lattice sites, and $\langle ...\rangle$ denotes that the sum goes over nearest neighbor pairs of sites $i$ and $j$. 
However, the B particles see a weaker disorder potential, and the parameters are such that the $B$ particles in isolation are in a thermal phase. The B system in isolation has a characteristic local bandwidth $\Delta$ and a dynamical timescale $\tau$. We couple the two systems together with a coupling Hamiltonian
\begin{equation}
H_{int} = \frac{g}{\sqrt{N}} \sum_{i, \alpha \leq N} c^{\dag}_i c_i d^{\dag}_{i, \alpha} d_{i, \alpha}
\end{equation}
%

The parameter $N$ controls the strength of back-action of the MBL system on the bath. In the limit of large $N$ (at constant $g$), back-action can be ignored; if we also take the bath to be at infinite temperature, as we discuss below, the bath can be treated as a classical external noise source. When $N$ is small and $n_B \lesssim n_A$, however, it is possible for the system to substantially alter the properties of the bath. 

In the latter sections of this paper, where we consider codimension one, we will restrict either the $A$ or the $B$ particles to live entirely on a `boundary layer' of the lattice. In the earlier part of the paper however (codimension zero), both $A$ and $B$ particles can live anywhere on the $d$ dimensional lattice. 

\section{MBL system coupled to classical noise}
\label{noise}
To simplify our analysis, we first discuss the case of large $N$ and infinite temperature, where back-action is negligible and the bath can therefore be regarded as a classical noise source. In this case, one can rewrite the system-bath interaction as $\sum_i c^\dagger_i c_i \phi_i(t)$, where $\phi_i(t)$ is a fluctuating classical field. Without loss of generality, we can take $\langle \phi_i \rangle = 0$, since any constant shift can be absorbed into $H_A$. In this approximation, the coupling constant $g$ can be absorbed into $\phi$, and determines the ``strength'' of the noise, $\Lambda^2 = \langle \phi^2(t) \rangle$. We note that when $W/\Lambda \ll 1$, then the noise strength is the largest energy scale in the A system. In this regime, the appropriate starting point is to account for the noise \emph{exactly} and treat the Hamiltonian of system A as a perturbation - see e.g. Ref.~\cite{alp}. As this situation does not fall within the framework of `MBL + bath' we will not discuss it further here, restricting our attention to situations where $\Lambda \ll W$. 

\begin{figure}
\includegraphics[width=\columnwidth]{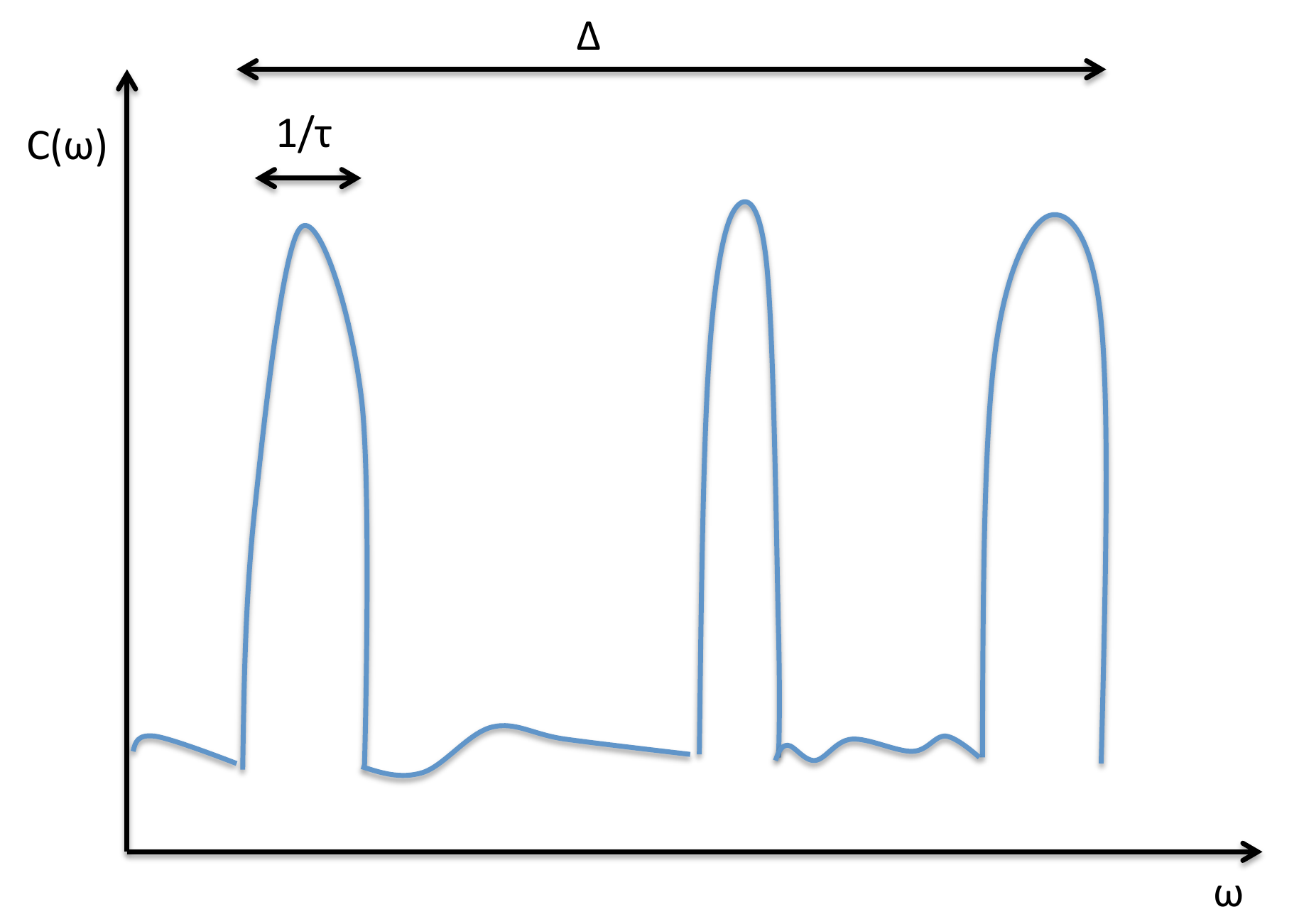}
\includegraphics[width = \columnwidth]{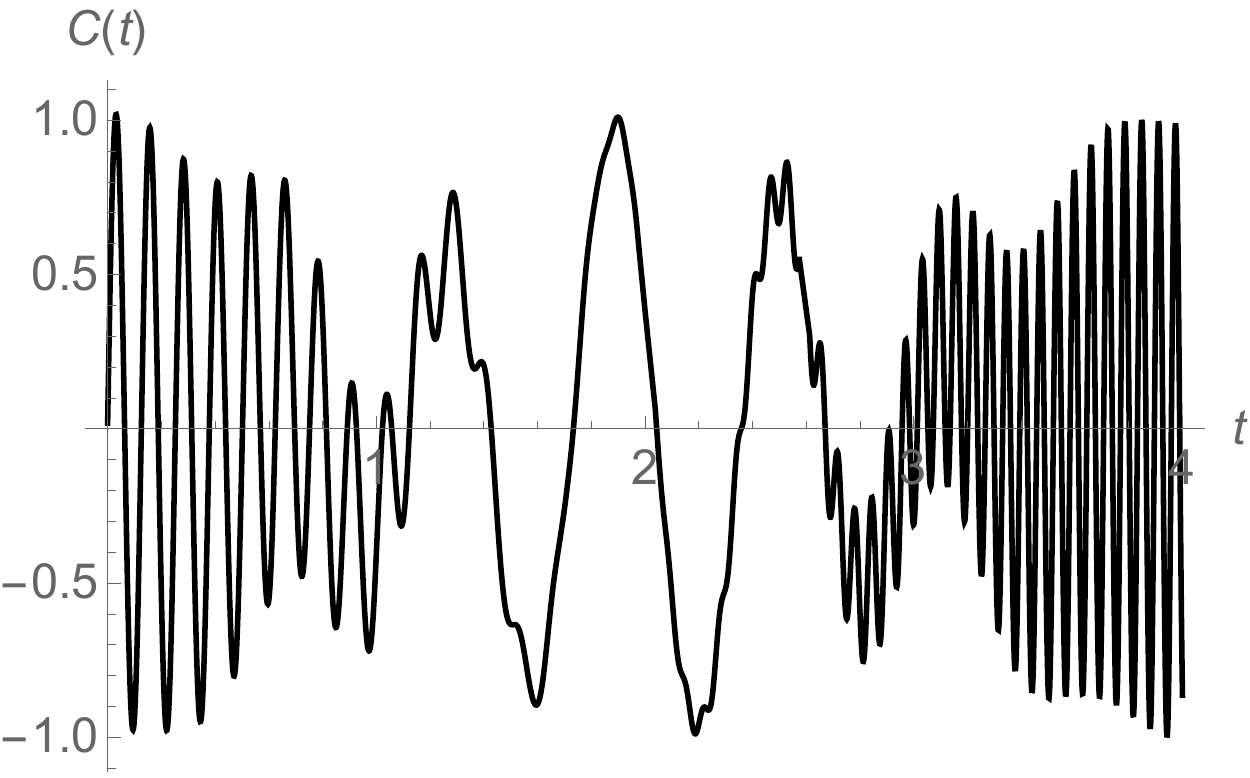}
\caption{\label{linewidth}Upper panel: An illustration of the spectral diffusion scenario. When the power spectrum of the noise is evaluated over a time window of order $1/\tau$ it may take the form above, with a collection of narrow spectral lines (of width $1/\tau$) spread over a frequency window of width $\Delta$. When the power spectrum is evaluated over a time window $\gg \tau$, these spectral lines will move around and the power spectrum will fill in everywhere in a window of width $\Delta$. The width of the power spectrum of the noise may thus depend on the timescale on which the noise is being probed, but will be lower bounded by $1/\tau$ and upper bounded by $\Delta$. Lower panel: a sample real-time noise trajectory in the spectral-diffusion regime, showing the separation of timescales between the rapid oscillations (with period $1/\Delta$) that change frequency on timescales $\tau$.}
\end{figure}

Apart from the strength, the properties of the noise that are most relevant for our purposes are its bandwidth $\tilde \Delta$ and its correlation time $\tilde \tau$. 
The bandwidth is defined as the frequency-space width of the power spectrum, viz. $C_i(\omega) = \int dt e^{i\omega t} \langle \phi_i(t) \phi_i(0) \rangle$.
In general, $\tilde \Delta$ and $1/\tilde \tau$ are distinct concepts, although $\tilde \Delta \agt 1/\tilde \tau$. A noise profile like that shown in Fig.\ref{linewidth}, corresponding, e.g., to a noise source with slow spectral diffusion, can have $\tilde \Delta \gg 1/\tilde \tau$ (in this situation, $\tilde \tau$ is a correlation time that is given by the decay time of the fourth-order correlator $\langle \phi^2(t) \phi^2(0) \rangle$). 

\begin{figure}
\includegraphics[width = \columnwidth]{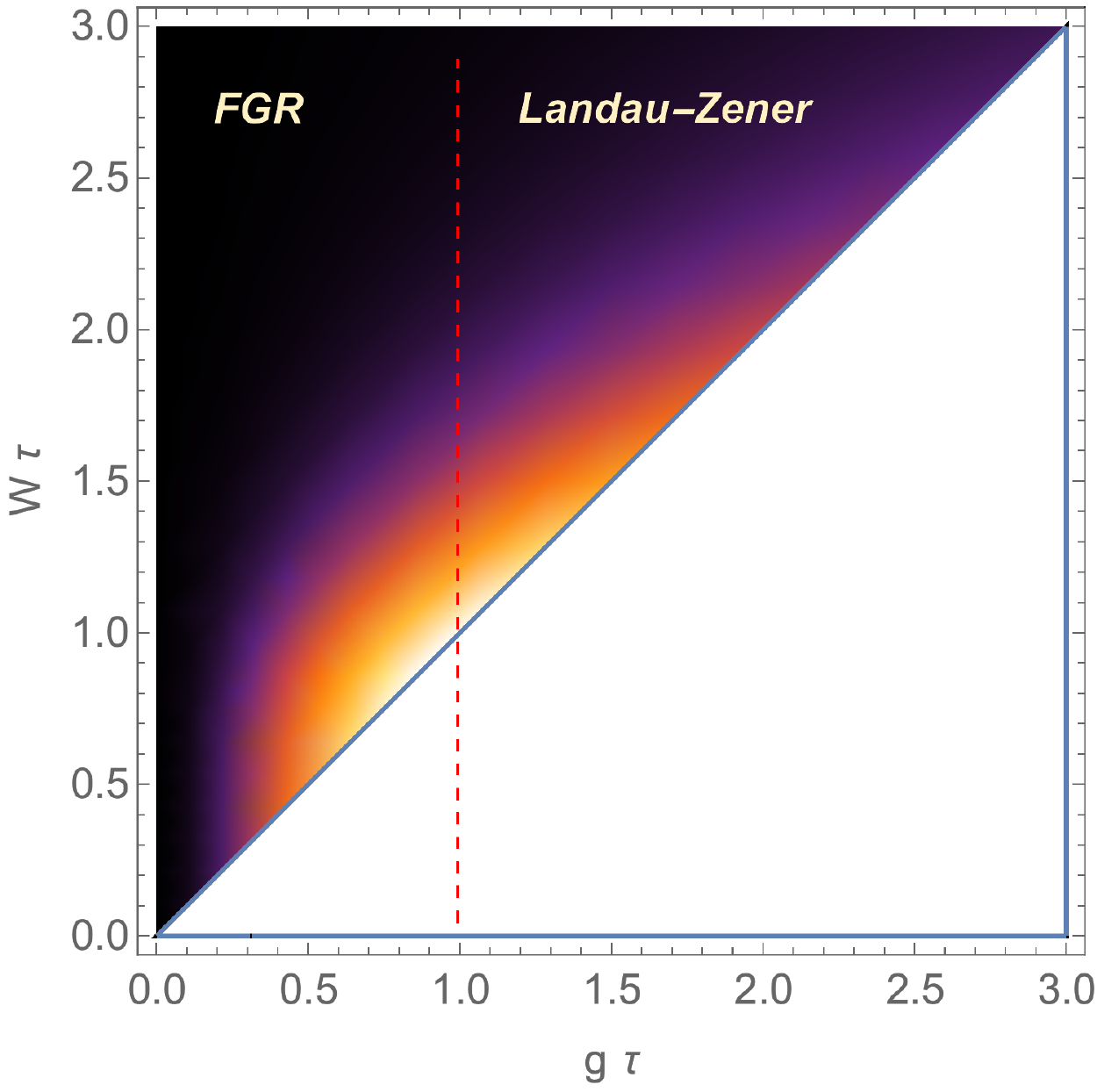}
\caption{\label{noisefig} Figure illustrating the parameter space of MBL + classical noise models, which is controlled by two parameters: $g \tau$ and $W \tau$. The color code is such that brighter colors correspond to faster relaxation. We restrict ourselves to the regime $W/g\gg 1$, since $W/g \ll 1$ does not fall into the framework of MBL+bath. The parameter $g \tau$ controls the general framework within which relaxation should be understood. When $g \tau \gg 1$ (but $W \gg g$) then relaxation is best understood in terms of Landau Zener transitions. When $g\tau \ll 1$ (and $W \gg g$) then relaxation is best understood in terms of the Golden Rule. In this latter case $W \tau$ controls the nature of the Golden Rule relaxation. For $W \tau \ll 1$ relaxation is dominated by the lowest order rearrangements, whereas when $W \tau \gg 1$ then relaxation is dominated by highly collective rearrangements (i.e. is bottlenecked by the small bandwidth of the bath). These different regimes are discussed in detail in the text.  }
\end{figure}

In the case of $\tilde \Delta \sim 1/\tilde \tau$, there are two dimensionless parameters governing the behavior of the system: these are the quantities $W \tilde \tau$ (which determines whether the noise is ``narrow-band'' or ``broad-band''), and $\Lambda \tilde \tau$ (which determines whether the Golden Rule is applicable). Thus there are three limiting behaviors consistent with our assumption that $ \Lambda \ll W$. These are (i) $\Lambda \ll W \ll \frac{1}{\tilde \tau}$, (ii) $\Lambda \ll \frac{1}{\tilde \tau} \ll W$, and (iii) $\frac{1}{\tilde \tau} < \Lambda < W$. We discuss all three in turn (see also Fig.\ref{noisefig} for a summary). 

{\bf The simplest limit is that of weak coupling to rapidly fluctuating noise: $\Lambda \ll W \ll 1/\tilde\tau$}. Here, Fermi's Golden Rule is evidently applicable and suggests a transition rate $\sim \Lambda^2 \tilde \tau$. This corresponds to an essentially Markovian bath, as considered in Ref.~\cite{NGH}. For an MBL system with exponentially decaying interactions, there is a log enhancement to Fermi's Golden Rule (\cite{NGH}) such that the `true' decay rate is 
\begin{equation}
\label{enhance}
\Gamma = \Lambda^2 \tilde \tau s \xi^d \ln^d \frac{W}{\Lambda^2 \tilde \tau}.
\end{equation} where $s$ is the entropy density i.e. a measure of the fraction of degrees of freedom that are `active.' This enhancement is obtained from the following line of reasoning: for a single degree of freedom coupled to classical noise, Fermi's Golden Rule predicts a decay rate $\Lambda^2 \tilde \tau$. For $N$ strongly coupled degrees of freedom, this decay rate is multipled by $N$, since any of the degrees of freedom can couple to the noise. For a system with interactions that fall off as $\exp(-r/\xi)$, two degrees of freedom should be considered strongly coupled if their mutual interaction $W \exp(-r/\xi)$ exceeds the decay rate. Self consistency then yields the expression above. Parenthetically, we note that for stretched exponential interactions $W \exp(-(r/\xi)^{\alpha})$, the logarithm is raised to a power $d/\alpha$, whereas for power law interactions $W (r/\xi)^{-\beta}$ one obtains a {\it power law} enhancement 
\begin{equation}
\Gamma = \Lambda^2 \tilde \tau s \xi^d \left(\frac{W}{\Lambda^2 \tilde \tau} \right)^{d/\beta}
\end{equation} 
(assuming, of course, that the exponent $\beta$ is large enough to be compatible with MBL \cite{Burin, yaodipoles}). All of the above expressions are accurate only when the `enhancement factor' is large compared to one (otherwise the relaxation rate is simply $\Lambda^2\tilde \tau$), and correspond to an inverse `$T_2$' time for the system (the `$T_1$' time is simply $1/\Lambda^2 \tilde \tau$ \cite{NGH}). 

{\bf A second simple limit is of very weak coupling to slowly fluctuating noise, i.e., $\Lambda \ll 1/\tilde \tau \ll W$}. This is the limit considered in Ref.~\cite{GN}; its key feature is that the frequency of a typical nearest-neighbor system transition greatly exceeds $1/\tilde \tau$. Again, the Golden Rule applies here, but it predicts that decay rates are strongly suppressed. 
The precise nature of the resulting behavior depends on the large-$\omega$ behavior of $C_i(\omega)$: if it falls off faster than $1/\omega^2$, relaxation is dominated by large-scale rearrangements and the relaxation rate is power law small in $1/\tilde \tau$, with a continuously varying exponent that depends on $W$ and temperature. If $C_i(\omega$) falls off as $1/\omega^2$ or slower, the dominant channel remains lowest-order, but has a suppressed rate $\sim \Lambda^2 C(W)$. In both cases, however, we expect the $\Lambda$-dependence of the decay rate to be quadratic (up to the enhancement factors associated with MBL \cite{NGH}, which take the same form as discussed above, but with $\Lambda^2 \tilde \tau$ replaced by the Golden Rule relaxation rate). 

{\bf The third and final limit falling within the framework of `MBL + noise' is $1/\tilde \tau \ll \Lambda \ll W$}. This is the regime of strong, slowly fluctuating noise. In the limit of $\tilde \tau \rightarrow \infty$, the reasoning of Ref.~\cite{nonlocal} suggests that the Golden Rule breaks down, and the dominant transitions are instead Landau-Zener crossings. In this case, the lifetime of the MBL system is dominated by the transition probability of its nearest adiabatic crossing. The distance to a Landau-Zener crossing is given~\cite{gkd} by $x \sim (1/s) \log(W/\Lambda)$, where $s$ is the entropy density, and the probability of an adiabatic crossing is $P_{ad} \sim \min(1, W^2 \tilde \tau \exp(-2x/\zeta)/\Lambda) \sim \min(1, (W^2/\Lambda) (\Lambda/W)^{2/(s\zeta)})$. The Golden Rule rate in this limit will be $\sim (1/\tilde \tau) P_{ad}$.

We now briefly consider the ``spectral-diffusion'' scenario in which $\Delta \gg 1/\tilde \tau$. We specialize to the case $\Delta \agt W$. In the Golden-Rule limit $\Lambda \tilde \tau \rightarrow 0$ we find that $\tilde \tau$ is irrelevant and the rates are given by substituting $\Delta$ for $1/\tilde \tau$ in the expressions above. This is because on the (very long) timescale associated with the decay, spectral diffusion allows the spectrum of the bath to `fill in' everywhere in a region of width $\Delta$. However, a crossover takes place when the resulting Golden-Rule rate becomes comparable to $1/\tilde \tau$. When the decay rate is comparable to (or faster than) $1/\tilde \tau$, the time-averaged bandwidth $\Delta$ is irrelevant to the physics: on these timescales the drive is close to monochromatic on each site, and does not ``sample'' over the bandwidth $\Delta$. Thus one expects the decay rate to be bottlenecked by $1/\tilde \tau$ in this regime, provided that $\Lambda \ll W$. (In the $\tilde \tau \rightarrow \infty$ limit, the physics corresponds to that of a system driven at a large number of incommensurate frequencies; whether such a system remains localized is at present an open question.) There is also an intermediate regime when spectral diffusion is partially but not wholly effective on the timescale set by the decay, and the relevant energy scale is intermediate between $1/\tilde \tau$ and $W$, and is in fact set self consistently by the decay rate. This regime was discussed at length in \cite{gnarxivv2}. 

\subsection{Quantum effects: discreteness and back-action}

\begin{figure}
\includegraphics[width=\columnwidth]{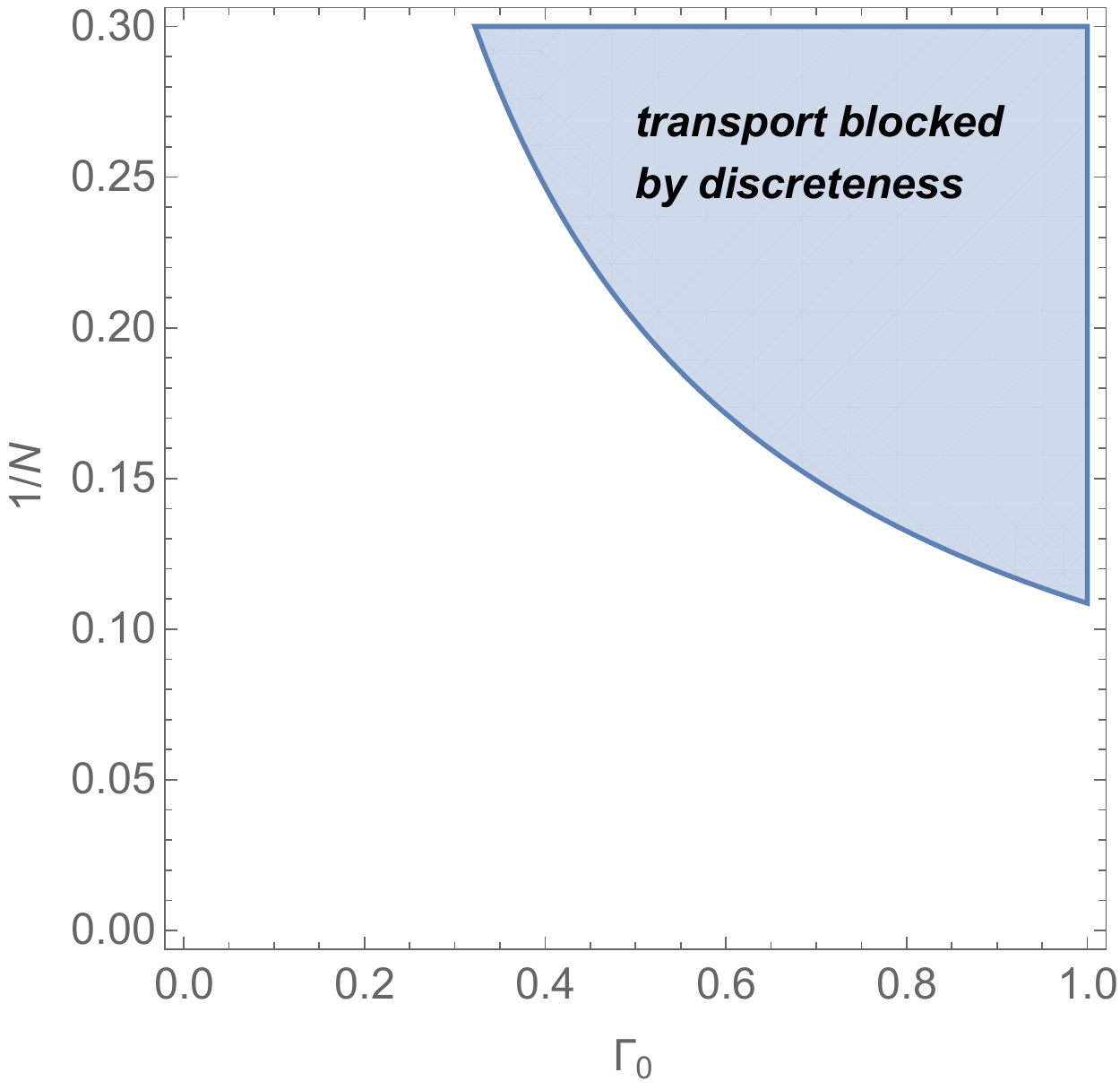}
\caption{\label{discreteness} Figure illustrating how, for not too large $N$ and not too weak noise, one can enter a regime where naive estimates of the relaxation rate violate the consistency criterion (\ref{consist}). In this regime relaxation is bottlenecked by discreteness of the accessible states in the bath. This regime is discussed at length in Sec.\ref{specialcases} and Sec.\ref{boundary}.}
\end{figure}

In the previous discussion we took the $N \rightarrow \infty$ limit, which allowed us to make two important simplifications. First, we were able to neglect the back-action of the system on the bath; and second, we were able to ignore possible complications related to the discreteness of the bath energy levels. These complications arise because of the following logic. Assuming the bath is in a thermal diffusive phase, entanglement spread takes place ballistically, so that on a timescale $t$ one can regard the bath as consisting of causally disconnected blocks of size $L_t \sim vt$. (In the subdiffusive Griffiths phase near the transition \cite{Kartiek, voskhusealtman, pvp}, $L_t \sim t^\alpha$ with $\alpha$ approaching zero at the MBL transition.) Thus, if a system level decays into the bath on a timescale $\Gamma$, it can only entangle with a region of size $L_{1/\Gamma}$ on this timescale. For the Golden Rule (or any other continuum approximation) to be consistent, we must require that $\Gamma$ exceed the level spacing of the bath on scale $L_{1/\Gamma}$, or in other words that

\begin{equation}
\label{consist}
\Gamma \agt \frac{\Delta v^d}{\Gamma^d} \exp[- s v^d/\Gamma^d].
\end{equation}
This condition is always satisfied at sufficiently weak coupling -- although this weak-coupling requirement becomes increasingly stringent as one approaches the MBL transition. Note that Eq.~\ref{consist} is a consistency condition: it is \emph{necessary} for Golden-Rule reasoning to apply but not sufficient. Note also that when there are $N$ flavors in the bath, $s \sim N$, so that in the large $N$ limit this consistency condition is automatically satisfied.See Fig.\ref{discreteness} for an illustration of these points. 

When this weak-coupling approximation fails, the leading interactions between the system and the bath are ``off-resonant'' processes such as Hartree and Stark shifts. Such shifts can in general {\it enhance} the effective disorder in the system, increasing it to $\sqrt{W^2+\Lambda^2}$, in effect {\it strengthening} localization (this is akin to the `Zeno localization' phenomenon discussed in \cite{HNPRS}). 

Another effect that is ignored when the bath is modeled as a classical noise source is the effect of {\it back action} on the bath. In particular, the coupling to the system causes the effective disorder strength in the bath to be increased. If the `bare' disorder strength in the bath is $w'$, then the disorder strength in the presence of coupling to the system becomes $\sqrt{(w')^2 + g^2/N}$ (adding scales in quadrature). In the limit $N\rightarrow \infty$ the back action is asymptotically weak and may be neglected, but at finite $N$, and particularly if the bath is only weakly ergodic, this increase in the effective disorder strength in the bath can be sufficient to drive the bath itself into a localized phase. This is the MBL proximity effect discussed in \cite{proximity}. 

\section{MBL+Bath: Fully quantum treatment}
\label{mblbath}
In this section we consider a fully quantum treatment of the `MBL+Bath' problem. 
The model we are considering was laid out in Sec.\ref{models}; we now enumerate the relevant energy scales and dimensionless parameters.
The dynamics of the $A$ system (which would, in isolation, be localized) is characterized by a characteristic local energy scale $W$ (which can generally be associated with the disorder bandwidth).
The dynamics of the $B$ system (which would, in isolation, be thermal) is characterized by a local energy bandwidth $\Delta$, a correlation time $\tau$, and an entropy density $s \propto N$.  Moreover, the spread of entanglement in the isolated $B$ system would proceed as $S \sim (t/\tau)^\alpha$, where $\alpha = 1$ in diffusive systems and $\alpha < 1$ in subdiffusive Griffiths phases. 
Deep in the thermal phase, $\alpha = 1$ and $\Delta \simeq 1/\tau$. In this section we restrict ourselves to the scenario $\Delta \sim 1/\tau$, deferring a discussion of the spectral diffusion scenario $\Delta \gg 1/\tau$ to the next section. 

It is helpful to rewrite these parameters as scale-dependent quantities. A block of linear size $L$ in the $B$ subsystem has a bandwidth $L^d \Delta$ [and correspondingly a many-body level spacing $L^d \Delta \exp(-s L^d)$], and becomes entangled on the timescale $t(L) \equiv L^{1/\alpha} \tau$. As discussed in the previous section, the relaxation rate determines a characteristic length-scale $L_\Gamma \sim (\Gamma \tau)^{-\alpha}$. This length-scale specifies a bandwidth, $s L^d_\Gamma \Delta$, as well as a level spacing
\begin{equation}
\label{discrete}
\delta_{\Gamma} \sim \min(\frac{1}{\tau}, \frac{s L_{\Gamma}^d}{\tau } \exp[- s L_{\Gamma}^d]).
\end{equation}
this energy scale will be an important point of reference in our analysis. We further denote by $\delta_g$ the level spacing associated with taking the Golden Rule result for the relaxation rate $\Gamma \sim g^2 \tau$. Finally, there is the coupling $g/\sqrt{N}$ between the two systems. 

Thus there are overall three independent dimensionful parameters $W, 1/\tau, g/\sqrt{N}$ and two dimensionless numbers $s$ and $N$ on which the physics may depend. Additionally there is an energy scale $\delta_{\Gamma}$ which is fully determined by the above parameters but is nonetheless important. Note that we have also assumed that the system and bath are interacting systems on the timescale set by $\Lambda$. When this is not true and system or bath are effectively non-interacting on the relevant timescale, then a different analysis must be used, and this is discussed in Appendix.\ref{noninteracting}. 

{\bf The dimensionless ratio  $W / g$ controls how strongly the A system is coupled to the B system}. When $W/g \ll 1$ then the coupling to the B system is the largest energy scale in the problem, and should be diagonalized first, before incorporating the Hamiltonian $H_A$ as a perturbation. This situation does not fall within the framework of `MBL+bath' and will not be discussed here. We will restrict our attention to $W/g \gg 1$.

{\bf The dimensionless ratio $\tau g /\sqrt{N}$ controls the strength of the back action on the B system}. When $\tau g/ \sqrt{N} \ll 1$ then the back action on the B system is weak, and when $\tau g/\sqrt{N} \gg 1$ the back action on the B system is strong. Note that $\tau g/\sqrt{N} \ll 1$ is a necessary (but not sufficient) condition for us to be able to model the B system as a classical noise source. Note also that $\tau g/\sqrt{N} \ll 1$ automatically guarantees $\delta_g \ll 1/\tau$, whereas in the strong back action regime $\delta_g \approx 1/\tau$, and there is little entanglement spreading in the bath on the timescale $t_g$. 

{\bf This straightaway allows us to identify {\bf $g \tau / \sqrt{N} \gg 1$} as a regime of strong back action}, where the bath cannot be modeled as a classical noise source. The coupling to the A system is the dominant energy scale for the B system, but the coupling is only a weak perturbation to the A system (since $W/g \gg 1$ by postulate) . In this scenario, the bath is likely localized by an `MBL proximity effect' \cite{proximity}. 

We henceforth specialize to the regime $g \tau /\sqrt{N} \ll1 $, when the back action on the bath is weak. Note that $g \tau / \sqrt{N} \ll 1$ automatically ensures $\delta_g < \Gamma_g$, so the discreteness of the bath is not an issue in this regime. 

The behavior in this regime is controlled by two parameters $g \tau$ and $W \tau$. 
{\bf The parameter $g \tau$ controls whether the B system is slowly or rapidly fluctuating on the timescale relevant for the A system.} This parameter is obtained by comparing the Golden Rule decay rate $g^2 \tau$ to the dynamical timescale in the bath. When $g \tau \ll 1$ the B system is rapidly fluctuating on the timescale relevant for the A system. Meanwhile, {\bf the parameter $W \tau $ controls whether we are in the broad or narrow band regimes}. $ W  \tau \ll 1$ is the broad band regime where the B system can easily supply enough energy to place rearrangements in the $A$ system on shell. In this limit the physics is essentially independent of $\tau$ (although not $\delta$). Meanwhile, $  W \tau \gg1$ is the narrow band regime where the relaxation rate (if the system delocalizes) is bottlenecked by $1/ \tau$. 

There are three limits compatible with our assumption $W/g \gg1 $ and $g \tau/\sqrt{N} \ll 1$, and these are (i) $W \tau \ll 1$ and $ g \tau \ll 1$, (ii) $W \tau \gg 1$ and $g \tau \gg 1$, and (iii) $W \tau \gg 1$ and $g \tau \ll 1$.  There are three distinct regimes which map onto the three models of MBL + classical noise discussed in Sec.\ref{noise}, namely



\begin{enumerate}

\item {\bf $ W\tau \ll $ and $g \tau \ll 1$}. This is the regime of a broad bandwidth bath where relaxation proceeds via the Golden Rule. Back action on the bath is weak and discreteness of the bath spectrum is unimportant so the bath can be modeled as a rapidly fluctuating classical noise source. In this limit, the bath generically delocalizes the system, and the behavior is as discussed in \cite{NGH} and \cite{JNB}, and also in Sec.\ref{noise} as the regime $\Lambda \ll W \ll 1/\tau$. 

\item {\bf $ W \tau \gg 1$  and $g \tau \ll 1$}. This is the regime of a good but narrow bandwidth bath that is able to place rearrangements in the A system on shell. Back action on the bath is weak and the bath can be modeled as a rapidly fluctuating classical noise source. In this limit, the bath delocalizes the system, but the relaxation rate is bottlenecked by $1/\tau$. This is the regime that was discussed in \cite{GN}, and also in Sec.\ref{noise} as the regime $\Lambda \ll 1/\tau \ll W$. 

\item {\bf $ W \tau \gg 1$  and $g \tau/\sqrt{N} \ll 1 \ll g \tau$}. In this limit, the bath delocalizes the system, but the dominant relaxation mechanism involves Landau Zener transitions rather than the Golden rule. This is the regime that was discussed in \cite{gkd}, and also in Sec.\ref{noise} as the regime $1/\tau \ll \Lambda \ll W$. 
\end{enumerate}

This concludes the survey of possibilities in the case when the bath is characterized by a single parameter and is not `protected' in any way. We have restricted ourselves to situations where the relevant energy scales are widely separated and the behavior can be straightforwardly deduced. Intermediate regimes where e.g. $g \tau/\sqrt{N} \approx 1$ are beyond the scope of the current analysis. Additionally, we have restricted ourselves to a regime where the A system is weakly coupled. When the A system is strongly coupled to the B system a different approach is called for, and either a localized or delocalized phase may result \cite{proximity}. 

\section{{Baths with multiple intrinsic timescales}}
\label{specialcases}
\label{spectral}
Thus far we have assumed that the B system is fully characterized by an entanglement spreading time $\tau$ and an associated energy scale $1/\tau$. However, as has been discussed in Sec.\ref{noise},  when the timescale on which the B system is being probed is long compared to $\tau$, spectral diffusion can lead to the emergence of a second energy scale $\mathcal{E}$, which is lower bounded by $1/\tau$, upper bounded by the local bandwidth of the B system $\Delta$, and is self consistently determined taking into account the relaxation rate in the system.  This makes a difference if we are in the regime $ 1/\tau \ll  W$  and $g \tau \ll 1$ when the bath is narrow bandwidth, rapidly fluctuating, and there is weak back action on the bath. In this case, the bottleneck on the relaxation becomes $\mathcal{E}$ instead of $1/\tau$ i.e. relaxation is faster than one would  naively expect. This situation was analysed in detail in \cite{gnarxivv2}. 

There are two generic situations when spectral diffusion is expected to be relevant. One is when the $B$ system is close to a localization transition. As the $B$ subsystem approaches its MBL transition, the timescale $\tau$ becomes much larger than $1/\Delta$, as the diffusion constant vanishes. In this regime, The structure of local spectral functions in the B system is as follows. A typical local operator has $\sim s$ spectral lines, in a bandwidth $\Delta$. Each spectral line has a characteristic ``width'' $\sim 1/\tau$, and decays exponentially or faster at frequencies $\gg \Delta$; this is implied, e.g., by rigorous results on absorption~\cite{ADH}. At intermediate timescales, the \emph{typical} spectral function is a Lorentzian (as one expects in the diffusive phase) or possibly a Levy-stable distribution (in the subdiffusive phase). 

The spectral diffusion scenario can also obtain in the regime where the bath is strongly coupled $g \tau /\sqrt{N} \gg 1$, but the bath is protected against localization because of symmetry, topology, or the existence of sufficiently long range interactions. 

{\bf In the case of protected baths, there is an additional {\it intermediate coupling regime} that can arise that has not been hitherto discussed. This is a regime where $g \tau/\sqrt{N} \gg 1$}(but $W/g \ll 1$). Even though this is a regime of strong back action on the bath, if the bath is protected against localization then the `proximity effect' is evaded. We can however enter a regime where $\Gamma_0 < \delta_{\Gamma_0}$, where $\Gamma_0$ is the relaxation rate determined from either the broad band Golden Rule, narrow band Golden Rule, or Landau Zener formulae, according to the relative sizes of $W$, $N$ $\tau$ and $g$. (This regime disappears in the large $N$ limit since $s \propto N$). In this regime, discreteness of the spectrum of the B system is important, and the bath thus cannot be modeled as a classical noise source. Instead, the B system enables relaxation in the A system by going to {\it high orders} in the coupling to the bath or by coupling to {\it highly collective rearrangements} in the system, which have a correspondingly smaller matrix element and relaxation rate.   This situation is analyzed in detail in Appendix \ref{bulk}. A key result is that in this regime the relaxation rate is bottlenecked by the discreteness of the bath spectrum and is effectively {\it independent} of $g$. The analysis in \cite{BanerjeeAltman} was in a similar regime, except that the analysis there was developed for a non-interacting system (such that the collective rearrangements were absent) and for a non-interacting bath (such that $\delta$ scales differently with $\Gamma$ to Eq.\ref{discrete}).

We now discuss these various types of `protected baths' that can arise, and comment briefly on each. 

\subsection{Topologically protected baths}
In Sec.\ref{mblbath} we assumed that the bath {\it could} get localized in a strong back action regime. However, if the bath is topologically protected against localization, then this conclusion must be revisited. Even though such a problem might naively be in the regime of strong back action on the bath (and weak coupling for the system), the end result must be delocalization of the composite. This scenario is relevant for e.g. the analysis in Ref.\cite{QHMBL}, where the bath in question was the (topologically protected) critically delocalized state at the center of a Landau level. Of course, the analysis in Ref.\cite{QHMBL} also differs in that the `bath' and system were not cleanly separated, as in Sec.\ref{models}, but were different parts of the same single particle spectrum. 

\subsection{Symmetry protected baths}
If the bath is protected against localization by a symmetry, then too delocalization of the composite system must result, even in the strong back action regime. One realization of such a scenario is when the bath consists of Goldstone modes. This scenario was discussed in e.g. Ref.\cite{BanerjeeAltman}, where the bath in question was the phonons in a Dyson chain. Another realization of a protected bath involves a bath made out of spin degrees of freedom for which the Hamiltonian has SU(2) symmetry, since such a system is also protected against localization \cite{vpp}. This latter realization was discussed in \cite{spinbath} in the context of the spin incoherent Luttinger liquid, where the charge and thermal transport properties in the presence of this spin bath were deduced. 

\subsection{Baths protected by long range interactions}
Systems with long range interactions that decay as power laws in space may support percolating networks of resonances \cite{Burin, yaodipoles}, that may act as a heat bath for the problem, triggering delocalization. If this does happen, then the heat bath in question will `live' on a sparse network of sites, and will be exceedingly narrow bandwidth. Transport in the presence of such a bath has unusual properties that have been explored in the low temperature limit in Ref.\cite{gpbgsm}. 

\section{MBL+Bath in codimension one}
\label{boundary}
The preceding discussion was for systems and baths coupled together with codimension zero i.e. the system and bath have the same dimensionality and are coupled everywhere in space. In this section we consider codimension one: a layer of thermal phase deposited on the surface of an MBL bulk, and a layer of MBL phase deposited on an thermal bulk. Note that this setup only makes sense if the bulk is in dimension $d > 1$, so that the boundary can itself be thermodynamically large (and hence capable of supporting either an MBL or thermal phase). 

\subsection{MBL Bulk with thermal boundary}
In this subsection we consider the behavior of an MBL system where a thermalizing quantum system is placed on its boundary.  Such a situation can be modeled within the framework outlined in Sec.\ref{models}, if the B particles are restricted to living on the boundary of the lattice on which the A particles live. What sort of behavior should we expect from such a setup? We note that a similar setup was analyzed in Ref.\cite{Chandran}, in the regime where the boundary bath was good (broad bandwidth, weak back action, rapidly fluctuating), and where the overall geometry was that of a $d$ dimensional cubic lattice of linear size $L$. We consider the generalized version of this problem, where we do not restrict the nature of the boundary bath or the system geometry. 

One possibility is that the `thermal' B system gets localized by the disorder coming from its coupling to the A system. Such a scenario may play out if the bath is in the `strong back action' regime $g \tau / \sqrt{N} \gg 1$, where $g$ is the coupling on the boundary and $\tau$ is the entanglement time for the bath. In this case, one simply has an MBL system with some extra localized degrees of freedom on the boundary. 

A more interesting possibility is that the boundary is in the weak coupling regime $g \tau/ \sqrt{N} \ll 1$, such that the B particles remain in a thermal phase, where they could in principle act as a heat bath for the A system. How should one then understand relaxation in the A system? We assume in the following that the A system is characterized by a single localization length $\xi_A$ (which we henceforth denote simply by $\xi$), ignoring the possible complications of multiple localization lengths \cite{HNO}. 

Recall that modes deep in the MBL bulk will be well localized, with exponentially small weight on the boundary. The matrix elements for coupling to the bath will thus fall off exponentially with distance from the boundary $g(r) \sim g \exp(-r/\xi)$. Meanwhile the parameters $W$, $\tau$, $s$ and $N$ are defined as previously, but $\delta$ is defined as 
\begin{equation}
\label{boundarydiscrete}
\delta_{\Gamma} \sim \min(\frac{1}{\tau}, \frac{s L_{\Gamma}^{d-1}}{\tau } \exp[- s L_{\Gamma}^{d-1}]).
\end{equation}
where, recall, $L_{\Gamma} = 1/\Gamma^{\alpha}(r).$ In the Golden Rule regime $\Gamma \sim g^2$ and $L_{\Gamma} \sim \exp(2\alpha r/\xi)$ such that $\delta_{\Gamma}$ decays as a {\it double exponential} function of $r$ as we go into the bulk. \footnote{One could argue that perhaps the degrees of freedom in the MBL system `on the way' to the boundary should be included in the size of the effective bath. However, this simply changes $L_{\Gamma}^{d-1}$ to $L_{\Gamma}^{d-1} \log L_{\Gamma}$ in the above formula, and does not significantly alter the behavior.  }

We note that back action will always be weak, since $g \tau / \sqrt{N} \ll 1$ at the boundary and since $g$ decays exponentially going into the bulk. We note also that $W/g \gg 1$ is ensured deep in the bulk, so the behavior deep in the bulk is guaranteed to fall into the `MBL + bath' formalism. Finally, $g \tau \ll 1$ deep in the bulk, so that the bath is rapidly fluctuating, and $\delta \ll g \ll W,\tau$. Deep in the bulk we are thus inevitably in the regime where the B system can be modeled as a rapidly fluctuating classical noise source, and the only question remaining is whether this noise source is narrow band or broad band. This last is determined by whether $W \tau \ll 1$ or $W\tau \gg 1$, and the relaxation rate is obtained from Fermi's Golden Rule, substituting $g(r)$ in for $g$ from Sec.\ref{noise}. Note that the relaxation rate will decline {\it exponentially} with distance from the boundary.  

Ref.\cite{Chandran} discussed the case when the boundary is {\it finite}, such that $L_{\Gamma}$ saturates to a maximum lengthscale $L$. In this case, $\delta$ saturates to a minimum value $\delta_{min} = \frac{L^{d-1}}{\tau} \exp(-s L^{d-1})$. There then emerges a depth $R$ beyond which $g(r>R) < \delta_{min}$, such that the Golden Rule becomes inapplicable and the relaxation rate drops to zero. This depth may be estimated as 
\begin{equation}
R \approx \xi s L^{d-1}
\end{equation}
At depths greater than $R$, the MBL system is effectively decoupled from the bath (no relaxation). Note however that this critical depth $R$ diverges as $L \rightarrow \infty$. 

This concludes our discussion of the dynamics deep in the bulk. We now consider the behavior of the $A$ system {\it near} the boundary. We assume that $W/g \gg 1$ even at the boundary, so that the A system can be described in the MBL + bath framework everywhere. If the boundary also satisfies $g \tau/\sqrt{N} \ll 1$, such that the back action on the bath is weak, then it can be readily verified that $\delta_{\Gamma} < \Gamma$ everywhere, so that the discreteness of the bath is also unimportant. In this case the effect of the bath can be modeled everywhere as classical noise. There are the usual three cases:

\begin{enumerate}

\item If $W \tau \ll 1$  then the bath can be everywhere modeled as a classical rapidly fluctuating broad bandwidth noise source, and the relaxation rate is given by the Golden Rule (with log enhancements as in Sec.\ref{noise} and with a matrix element that decays as $\exp(-r/\xi)$). 

\item If $W \tau \gg 1$ and $g \tau \ll 1$ then the relaxation rate is given by the Golden Rule as before, but bottlenecked by $1/\tau$, and with a matrix element that decays as $\exp(-r/\xi)$.

\item If $W\tau  \gg 1$ and $g(0) \tau/\sqrt{N}\ll 1\ll g(0)\tau $, then {\bf there is a crossover behavior as a function of depth}. Deep in the bulk $g(r) \tau \ll1$ such that the noise source is rapidly fluctuating and relaxation is described by the Golden Rule, whereas close to the boundary $g(r) \tau \gg 1$ such that the noise source is slowly fluctuating and relaxation is described in terms of Landau Zener transitions. The crossover between the two pictures is at a radius $r_c \approx \xi \ln (g \tau)$. In both regimes the relaxation rate decreases exponentially with distance from the boundary, but with different decay lengths. 

\end{enumerate}

{\bf A different behavior can arise when the bath is protected against localization, and we are in the strong back action regime $g(0) \tau /\sqrt{N} \gg 1$}. Even though the bath is protected against localization, the relaxation rate obtained from the Golden Rule (or Landau Zener) calculation can violate Eq.\ref{consist}, near the boundary, such that in the near boundary regime the discreteness of the bath becomes important. In this event the bath cannot be modeled as a classical noise source close to the boundary (although it can be so modeled deep in the bulk). Relaxation close to the boundary then requires going to high orders in the coupling $g$, or making use of highly collective rearrangements that couple only weakly to the boundary bath. This scenario is analyzed in detail in Appendix \ref{boundaryapp}, and leads to a relaxation rate that is bottlenecked by discreteness of the bath spectrum, and is not only independent of $g$ but saturates to a constant for distances $r < r_c$ from the boundary. For distances $r > r_c$ of course the relaxation rate continues to decay exponentially with distance from the boundary. 

The final possibility (which we do not discuss since it does not fall into the MBL+bath framework), is that close to the boundary the system is strongly coupled to the bath, $W/g \ll 1$, and the MBL+bath framework only starts to apply at depths greater than $\xi \ln W/g$. This concludes our survey of MBL systems coupled to boundary baths. 

\subsection{Thermal phase with MBL boundary}
We now consider the situation where the B particles live on a $d>1$ dimensional lattice, and the $A$ particles are restricted to the $d-1$ dimensional boundary. If $g\tau/\sqrt{N} \ll 1$ at the boundary (weak back action at the boundary), then the MBL system simply gets delocalized, and falls into the appropriate class discussed in Sec.\ref{mblbath} according to the boundary values of the relevant parameters. A more interesting regime is when $g \tau / \sqrt{N} \gg 1$ at the boundary but $W/g \gg 1$. This further implies that at the boundary $W \gg 1/\tau$. This situation still falls into the MBL + bath framework, but if the B system consisted of only the boundary layer, the end result would be that the bath ends up localized by `proximity effect.'  However, the B system lives in a higher dimensional space to the A system, and the coupling to the A system will fall off as we go deep into the bulk, such that sufficiently far into the bulk we will be in the weak back action regime, and will end up with a bath capable of delocalizing the MBL boundary. We are assuming here that the B system {\it does} see disorder so that the correlation length $r_0$ in the B system is finite. 

If we assume that the coupling falls off as $g (r/r_0)^{-\chi}$ with distance from the boundary, then the shell of depth $r_0 (g \tau)^{1/\chi} > r > r_0 (g \tau /\sqrt{N})^{1/\chi}$ will constitute a slowly fluctuating bath that will lead to relaxation in the MBL boundary due to Landau Zener transitions, with coupling $\Lambda \approx \sqrt{N}/\tau$, whereas the region at depth   $r > r_0 (g \tau)^{1/\chi}$ will constitute a rapidly fluctuating bath that will lead to relaxation via the Golden Rule, with coupling $\Lambda \approx 1/\tau$. These two channels should be added in quadrature to determine the relaxation rate in the A system, which will be power law small in large $\tau$, but will be independent of $g$. 

We conclude by speculating as to the possibilities for a stronger result. If the `bath' were extremely weak, right on the cusp of a localization transition, could the application of an MBL boundary layer trigger a `localization avalanche' whereby the whole system gets localized? Such a scenario may play out as follows: the coupling to the A system causes the boundary layer to become frozen. The next layer then sees additional disorder coming from the boundary layer and freezes in turn, itself constituting a source of disorder for the third layer, and so on. If the localization transition is indeed second order (as is widely believed), then this scenario seems highly unlikely, since the application of an MBL boundary layer will not alter the disorder strength in the bulk, and the disorder strength is (presumably) the parameter driving the transition. However, if the MBL transition were {\it first} order, so that there was a co-existence regime, and the B system happened to be in the thermal phase in the co-existence regime, then indeed applying an appropriate boundary condition could trigger a localization avalanche causing the entire bulk to localize. We note too that recent numerical results \cite{Luitz} appear to support a scenario where the localization transition has at least some first order character, in that the local entropy density appears to show a discontinuity. Since we are not aware of any arguments conclusively establishing that the MBL transition must be second order, we cannot exclude the possibility of such a `localization avalanche,' which would seem to be an interesting topic for future work.

\section{Conclusions and open questions}
\label{conclusions}
We have discussed the behavior of an MBL system weakly coupled to a bath. When the back action on the bath is weak and the discreteness of the bath levels unimportant then the bath can be modeled as a classical noise source. In Sec.\ref{noise} we discussed the behavior of an MBL system coupled to a classical noise source, and highlighted three different regimes of relaxation. In Sec.\ref{mblbath} we introduced a fully quantum treatment of an MBL system weakly coupled to a heat bath. We identified a small number of parameters that control the physics, and discussed limiting regimes where these parameters were widely separated. There turned out to be four distinct regimes: three corresponding to the three distinct relaxation regimes for an MBL system subjected to classical noise, and an intrinsically quantum regime of {\it strong back action}, where the bath can itself get localized by the disorder coming from the MBL system. In Sec.\ref{specialcases} we discussed the spectral diffusion scenario where the inverse correlation time of the bath is much less than the local bandwidth. This scenario obtains in baths close to the localization transition and baths protected against localization by topological or symmetry considerations, or long range interactions. In this case there is also an additional intrinsically quantum regime that can arise. This is a regime where the discreteness of the bath energy levels is important, such that the dominant relaxation mechanisms involve high order coupling or highly collective rearrangements, and where the relaxation rate is independent of the coupling $g$. Finally, in Sec.\ref{boundary} we discussed the behavior of the `codimension one' problem, where the system and bath do not have the same dimensionality. We discussed first the case of an MBL system with a boundary bath. When the bath is unprotected and in the weak back action regime there arise the usual three distinct regimes of effectively classical noise. However, one of these three cases involves a crossover between different models of noise as a function of distance from the boundary. When the bath is protected against localization and in the strong back action regime there arises an intrinsically quantum regime where discreteness of the bath matters close to the boundary and leads to a relaxation rate that is depth independent in the near boundary regime. We also discussed the case of a thermal system with an MBL boundary, and speculated as to the possibilities for a localization avalanche. We trust that the framework introduced in this paper will prove useful for future investigations of MBL systems coupled to thermalizing environments. 

{\bf Acknowledgements:} We acknowledge useful conversations with P.W. Anderson, Ravin Bhatt, Anushya Chandran, Eugene Demler, Sonika Johri, Vedika Khemani, Michael Knap, Andrew Potter, Antonello Scardicchio, S.L. Sondhi and especially David A. Huse. 
\appendix
\section{Noninteracting limits}
\label{noninteracting}
 In the main text we assumed that both system and bath are effectively interacting on the relevant timescales. If the system is effectively non-interacting on the timescale $t_g$ then the problem maps onto the well studied problem of a {\it non-interacting} Anderson insulator coupled to a bath. There are then numerous differences to the preceding analysis. For example, there are no `enhancement factors' analogous to Eq.\ref{enhance}, and there is no scope for collective rearrangements in the narrow band regime. Instead the dominant paradigms are variable range hopping and the Golden Rule. 
 
If the bath is effectively non-interacting on the relevant timescale, then the main difference pertains to the scaling of $\delta_{\Gamma}$, which is modified from Eq.\ref{discrete} to $\delta \sim \frac{1}{\tau s L_{\Gamma}^d}.$ This is the case in e.g. the analysis of Ref.\cite{BanerjeeAltman}. The absence of interactions in the bath also makes it easier for the coupling to localize the bath, particularly in low dimensions, see e.g. the discussion of weak localization in Ref. \cite{proximity}. 

\section{\bf Discrete bath spectrum, codimension zero}
\label{bulk}
In this appendix we discuss the codimension zero problem with $W/g \gg 1$ and $g \tau/\sqrt{N} \gg 1$ where the bath is protected against localization, but where the Golden Rule (or Landau Zener) calculation yields a relaxation rate $\Gamma_0$ such that $\Gamma_0 \ll \delta_{\Gamma_0}$, violating the consistency condition Eq.\ref{consist}. In this case the relaxation rate is bottlenecked by the discreteness of the bath spectrum. Relaxation must occur through either high order coupling to the bath, or through collective processes, both of which have a smaller matrix element and hence a smaller relaxation rate satisfying $\Gamma \ge \delta_{\Gamma}$. Thus with logarithmic accuracy we find that $\Gamma$ is the solution to the self consistency equation
\begin{equation}
 \label{matrixelementsaturateszero}
 \Gamma = \frac{1}{\tau}  [-\ln (\Gamma \tau)/s]^{-1/d\alpha} 
 \end{equation}
 Note that this equation is {\it independent} of $g$ and thus in this regime the relaxation rate is independent of the coupling between system and bath. This is however a regime of intermediate $g$. At smallest $g$ one has $\Gamma_0 \ge \delta_{\Gamma_0}$ and returns to the model of classical noise. 
 

%
%

\section{\bf Discrete bath spectrum, codimension one}
\label{boundaryapp}

Here we discuss the codimension one problem where a  $d-1$ dimensional thermal system which is protected against localization is placed on the boundary of a $d$ dimensional MBL system. We are working in the regime where $g(r) = g(0) \exp(-r/\xi)$, $W/g(0) \gg 1$ and $g(0) \tau /\sqrt{N} \gg 1$, when the Golden Rule (or Landau Zener) calculations predict a relaxation rate $\Gamma_0$ violating the consistency condition $\Gamma_0 \ge \delta_{\Gamma_0}$, where $\delta_{\Gamma_0}$ is given by Eq.\ref{boundarydiscrete}. In this case, relaxation near the boundary will involve high order couplings to the bath or high order rearrangements in the system. Either way, the relaxation rate will be bottlenecked by the discreteness of the bath spectrum and will be $\Gamma \ll \Gamma_0$, satisfying $\delta_{\Gamma} = \Gamma$. Thus with logarithmic accuracy we get $L_\Gamma \approx [-\ln (\Gamma\tau)/s]^{1/(d-1)}$ and 
\begin{equation}
 \label{matrixelementsaturates}
 \Gamma = \frac{1}{\tau}  [-\ln (\Gamma \tau)/s]^{-1/(d-1)\alpha}
\end{equation}
Note that this equation is independent of $g$ (and hence distance from the boundary) and thus the relaxation rate is {\it constant} in the near boundary regime. The equation \label{matrixelementsaturates} only applies however when the relaxation rate predicted by the standard Golden Rule calculation violates the consistency condition. Sufficiently far from the boundary (at $r > r_c$), we inevitably satisfy $\Gamma_0 \ge \delta_{\Gamma_0}$, since $\Gamma_0$ declines exponentially with distance from the boundary and $\delta_{\Gamma_0}$ declines exponentially with $\Gamma_0$. Far from the boundary therefore we return to Golden Rule behavior with a decay rate the falls off exponentially with distance from the boundary. The critical distance $r_c$ can be estimated by setting $\Gamma_{g(r_c)} = \delta_{\Gamma_{g(r_c)}}$. For the broad bandwidth case where the golden rule predicts a relaxation rate $g^2 \tau$ this yields
\begin{equation}
r_c \approx \frac{\xi}{2 \alpha (d-1)} \ln \left(\frac{- 2 \ln g \tau}{s (\tau g)^{-\alpha (d-1)}} \right) \nonumber
\end{equation}

\centering
\bibliography{General_bib}
\bibliographystyle{careerbibstyle}

\end{document}